\documentclass[twocolumn,showpacs,preprintnumbers,amsmath,amssymb,superscriptaddress,prl,aps]{revtex4}
\def\cs{$^{137}$Cs~}

\def\Am{$^{241}$Am~}

\def\naitl{NaI(T$\ell$)~}
\def\csitl{CsI(T$\ell$)~}
   
\def\Journal#1#2#3#4{{#1} {\bf #2}, #3 (#4)}
\def\NIM{Nucl. Instrum. Methods}

\def\PLB{Phys. Lett.  B}
\def\ASP{Astropart. Phys.}
\def\PRL{Phys. Rev. Lett.}

\def\PRD{Phys. Rev. D}

\def\NJP{New J. Phys.}
\def\etal{{\it et al.}}

\usepackage{dcolumn}
\usepackage{bm}
\usepackage{graphicx}
\usepackage{amsmath}
\usepackage{latexsym}
\usepackage{amsfonts}
\usepackage{amssymb}
\usepackage{array}
\usepackage{epsfig}
\usepackage{multirow}

\begin{document}

\title{New Limits on Interactions between Weakly Interacting Massive Particles and Nucleons Obtained with CsI(Tl) Crystal Detectors
 }

\author{S.C.~Kim}
\affiliation{Department of Physics and Astronomy, Seoul National University, Seoul, 151-747, Korea}

\author{H.~Bhang}
\affiliation{Department of Physics and Astronomy, Seoul National University, Seoul, 151-747, Korea}

\author{ J.H.~Choi}
\affiliation{Department of Physics and Astronomy, Seoul National University, Seoul, 151-747, Korea}

\author{W.G.~Kang}
\affiliation{Department of Physics, Sejong University, Seoul, 143-747, Korea}

\author{B.H.~Kim}
\affiliation{Department of Physics and Astronomy, Seoul National University, Seoul, 151-747, Korea}

\author{H.J.~Kim}
\affiliation{Department of Physics, Kyungpook National University, Daegu, 702-701, Korea}

\author{ K.W.~Kim}
\affiliation{Department of Physics and Astronomy, Seoul National University, Seoul, 151-747, Korea}

\author{S.K.~Kim}
\email[]{skkim@hep1.snu.ac.kr}
\affiliation{Department of Physics and Astronomy, Seoul National University, Seoul, 151-747, Korea}

\author{Y.D.~Kim}
\affiliation{Department of Physics, Sejong University, Seoul, 143-747, Korea}

\author{ J.~Lee}
\affiliation{Department of Physics and Astronomy, Seoul National University, Seoul, 151-747, Korea}

\author{ J.H.~Lee}
\affiliation{Department of Physics and Astronomy, Seoul National University, Seoul, 151-747, Korea}

\author{ J.K.~Lee}
\affiliation{Department of Physics and Astronomy, Seoul National University, Seoul, 151-747, Korea}

\author{ M.J.~Lee}
\affiliation{Department of Physics and Astronomy, Seoul National University, Seoul, 151-747, Korea}

\author{ S.J.~Lee}
\affiliation{Department of Physics and Astronomy, Seoul National University, Seoul, 151-747, Korea}

\author{ J.~Li}
\affiliation{Department of Physics and Astronomy, Seoul National University, Seoul, 151-747, Korea}

\author{ J.~Li}
\affiliation{Department of Engineering Physics, Tsinghua University, Beijing, 100084, China}

\author{ X.R.~Li}
\affiliation{Department of Physics and Astronomy, Seoul National University, Seoul, 151-747, Korea}

\author{ Y.J.~Li}
\affiliation{Department of Engineering Physics, Tsinghua University, Beijing, 100084, China}

\author{ S.S.~Myung}
\affiliation{Department of Physics and Astronomy, Seoul National University, Seoul, 151-747, Korea}

\author{ S.L.~Olsen}
\affiliation{Department of Physics and Astronomy, Seoul National University, Seoul, 151-747, Korea}

\author{ S.~Ryu}
\affiliation{Department of Physics and Astronomy, Seoul National University, Seoul, 151-747, Korea}

\author{ I.S.~Seong}
\affiliation{Department of Physics and Astronomy, Seoul National University, Seoul, 151-747, Korea}

\author{J.H.~So}
\affiliation{Department of Physics, Kyungpook National University, Daegu, 702-701, Korea}

\author{ Q.~Yue}
\affiliation{Department of Engineering Physics, Tsinghua University, Beijing, 100084, China}

\collaboration{KIMS Collaboration}
\noaffiliation

\date{\today}

\begin{abstract}
New limits are presented on the cross section for Weakly Interacting Massive Particle (WIMP) nucleon scattering 
in the KIMS \csitl detector array at the Yangyang Underground Laboratory. 
The exposure used for these results is 24524.3 kg$ \cdot $days.
Nuclei recoiling from WIMP interactions are identified by a pulse shape 
discrimination method. A low energy background due to alpha emitters on the crystal
surfaces is identified and taken into account in the analysis.
The detected numbers of nuclear recoils are consistent with zero  
and 90\% confidence level upper limits on the WIMP interaction rates are set for
electron equivalent energies from 3 keV to 11 keV.  
The 90\% upper limit of NR event rate for 3.6-5.8 keV
corresponding to 2-4 keV in \naitl is 0.0098 counts/kg/keV/day which is below the
annual modulation amplitude reported by DAMA.   
This is incompatible with interpretations that enhance the 
modulation amplitude such as inelastic dark matter models. 
We establish the most stringent cross section limits on spin-dependent WIMP-proton elastic scattering
for the WIMP masses greater than 20 GeV/$c^2$. 
\end{abstract}

\maketitle

\setlength\arraycolsep{1pt}

Astronomical observations have led to the conclusion that
the majority of the matter in our universe is invisible, 
exotic and non-relativistic dark matter~\cite{wmap_7year}. However, 
the identity of the dark matter is still not known.
One possible source are WIMPs, which are candidates for particle dark matter
that naturally occur in theories that extend
the standard model of the particle physics for
reasons independent of the dark matter problem~\cite{Bertone}.
There have been a number of experiments that search for WIMPs in our galaxy by looking 
for nuclei recoiling from WIMP-nucleus scattering
~\cite{skkim10,paper_dama_libra,prllee,ana_naiad2005,cresst2,edelweiss2,xeplin3,xenon100,cdms2,picasso,coupp}.
To date, there are several experiments that interpret their results as being possibly due to WIMP signals including DAMA, CoGeNT and CRESST~\cite{paper_dama_libra,cogent,cresst_new}. 
In particular, the DAMA results have attracted considerable attention since it has  
reported observations of an annual modulation of WIMP-like signals with a claimed significance of 9 $\sigma$~\cite{paper_dama_libra}.
This has spurred a continuing debate concerning the observation of WIMPs that has lasted for over a decade.
WIMP-nucleon cross sections inferred from the DAMA modulation are
in conflict with limits from other experiments
that directly measure nuclear recoils~\cite{prllee,ana_naiad2005,cresst2,edelweiss2,xeplin3,xenon100,cdms2,picasso,coupp,savage}.
In attempts to reconcile these results, various schemes have been suggested,
including the inelastic dark matter (iDM) model~\cite{idm_chang}, in which an
excited state of the dark matter particle is hypothesized and the dominant
WIMP-nucleon scattering process involves a transition to this excited state.
Recently, strong constraints on the allowed iDM model parameter space have been reported~\cite{cdms2,idm_kai,zeplin_idm,xenon_idm}. 

The KIMS (Korea Invisible Matter Search) collaboration is performing direct
searches for WIMPs using \csitl detectors in the Yangyang 
Underground Laboratory (Y2L).  \csitl is a commonly used scintillating crystal
with $^{133}$Cs~ and $^{127}$I~ target elements that are sensitive to both 
spin-independent (SI) and spin-dependent (SD) interactions~\cite{skkim10}. 
Furthermore,  a pulse shape discrimination (PSD) technique makes it possible
to distinguish nuclear recoil (NR)-induced signals from electron recoil 
(ER)-induced signals on a statistical basis.  Results from the KIMS experiment
based on a four-crystal array are reported in Ref.~\cite{prllee}.  The detector 
has been upgraded to a 3$\times$4 \csitl crystal array with total mass 103.4 kg. 
Each detection module consists of a low-background \csitl crystal ($ 8 \times 8 \times 30 $ cm$^3$)~\cite{csicrystal} with photomultiplier tubes (PMTs)
mounted at each end. The data were collected between September 2009 and August 2010.
The crystal array is surrounded by a shield consisting of:
10~cm of copper; 5~cm of polyethylene; 15~cm of lead; and 30~cm of 
liquid-scintillator-loaded mineral oil to stop external neutrons and gammas and
veto cosmic-ray muons.  The trigger condition is two or more photoelectrons (PEs)
in each PMT within a 2\,$\mu $s time window.  Amplified PMT signals 
are digitized by 400 MHz flash analog-to-digital converters (FADC); 
the total recorded time window for an event is 40\,$\mu$s, of which 25\,$\mu$s
is analyzed.   Since high energy muon interactions in the \csitl crystal can produce a long tail 
that may last for several tens of milliseconds,  we veto events that occur
less than 50~milliseconds after a muon coincidence with the \csitl detector in the
off-line analysis.  The muon coincidence rate is 6-7 per hour and the dead time from this
veto is negligible.  In the energy region below 10~keV, PMT noise produces a serious background.  
To characterize PMT noise-induced events,  a PMT Dummy Detector (PDD) module that has the same 
structure as a \csitl detector module with the crystal replaced by a clean, transparent and empty
acrylic box is included in the shielded volume and operated  simultaneously with the \csitl detector array.
The event selection  efficiency is detector-module dependent and ranges from 20$\sim$40\% at 3~keV
to 40$\sim$70\% at 10~keV.   Events that trigger two or more detector modules in the array are
rejected off-line since a WIMP interacts only with the one nuclei in a detector: a NR event that is
confined to a single detector module is the experimental signature for a WIMP-nucleus scattering.

We determine the NR event rate from a PSD analysis.  To characterize the PSD, we use
the quantity we call LMT10,  which is the logarithm of the mean time of an event calculated
over a 10 $\mu$s interval that starts with the first detected PE.  Specifically, 
 MT10 (mean time over a 10 $\mu$s)$= \frac{\sum_{t_i<10\mu s}{A_i\times t_i}}{\sum_{t_i<10\mu s}{A_i}}$, where 
$A_i$ is the area of the $i^{\rm th}$ cluster,  which is usually equivalent to a single PE, of an event.

\begin{figure}[!htp]
\centerline{\scalebox{1.0}{\includegraphics[width=0.5\textwidth]{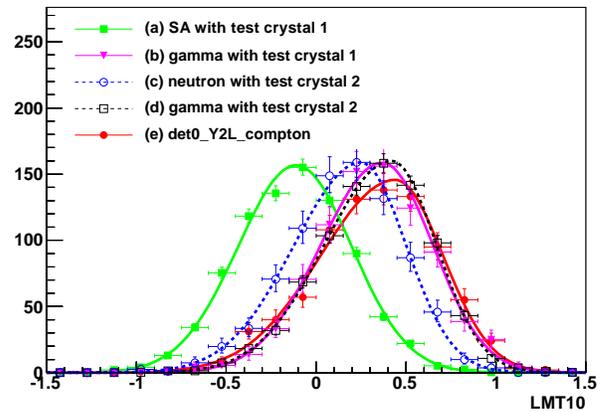}}}
\caption{LMT10 distributions at 3 keV for (a) SA with test crystal 1,(b) gammas with test crystal 1, (c) neutrons with test crystal 2, 
 (d) gammas with test crystal 2 (e) Compton scattering events in detector~0
used in the WIMP search. This data sample is obtained at $(25.2\pm 0.2)^{\circ}\mathrm{C}$ for comparisons
with other reference data. 
\vspace{-15pt}
} 
 \label{complogrmt10b}
\end{figure}

We determine the expected LMT10 distribution for ER events by irradiating small \csitl crystals ($3 \times 3 \times 1.4$ cm$^3$,
test crystals 1 \& 2) with gamma rays from a \cs radioactive source and also from Compton scattering events
detected in the array. 
The test crystals are small pieces that have been cut from the same ingots from which crystals used in WIMP search were also cut. 
 The expected distributions for NR events were determined
by exposing test crystal 2 to neutrons from a Am-Be source.
With the higher sensitivity of the current exposure, we detected a 
previously unseen third component with a mean time distribution that is faster than that for NR events.   
Studies show that this originates from alpha decays of radioactive isotopes
that adhere to the crystal surfaces, {\it i.e.,} surface alphas (SA), with charactersitics
described in detail in Ref.~\cite{SApaper}. 
To determine the LMT10 distribution for SA events, we contaminated test crystal 1 with Rn progenies,
and studied its response for events tagged as outgoing alphas from the crystal surface~\cite{SApaper}.  
Figure~\ref{complogrmt10b} shows a comparison of the three different reference LMT10 distributions.

The test-crystal measurements were done at temperatures of $(25.4\pm 0.3)^{\circ}\mathrm{C}$ for SA 
and $(25.3\pm 0.7)^{\circ}\mathrm{C}$ for NR.  
Throughout the WIMP search data-taking period, the crystals were kept in the $20 \sim 21.6^{\circ}\mathrm{C}$ 
temperature range, with variations that depend on the detector position; each detector
had rms temperature fluctuations of $\sim 0.2^{\circ}\mathrm{C}$. 
The mean values of the ER LMT10 distributions of the 12 detectors determined {\it in situ} have an average value of 
$0.62$ with  an $rms$ spread of $0.035$ for 59.54~keV gammas from an \Am source.
Since the temperature conditions of the reference samples used to determine the NR and SA distributions
are different,  some adjustment is necessary.
Previous studies~\cite{ana_naiad, exp_hslee} have shown that the ratio 
$R_{\tau}=\tau_{n}/\tau_{e}$ is independent of temperature,
where $\tau_{n}$ is the mean time of NR and $\tau_{e}$, the mean time of ER. By comparing the LMT10 distribution 
from gamma irradiation on a test crystal and that of the crystals used in the WIMP search,
we adjust the LMT10 distribution of NR events and SA events. 
We assume that there are only three components --NR, ER, and SA-- in the WIMP search data and
form the logarithm of the likelihood function for each energy bin as: 
\begin{eqnarray*}
        F_{i}&=&-\log(\mathcal{L}) \qquad \qquad \qquad \qquad \qquad \qquad \qquad \qquad \quad \quad\\
         &=& -\sum_{k=1}^{n_i}\log(  f_{N,i}P_{N,i}(x_k)+f_{S,i}P_{S,i}(x_k) \\
         & & \qquad \qquad  +(1-f_{N,i}-f_{S,i})P_{E,i}(x_k)),
\end{eqnarray*}
where $x_k$ is the LMT10 of the $k^{\rm th}$ event, the index $i$ denotes the $i^{\rm th}$ energy bin, $n_i$ is the number of events in the $i^{\rm th}$ energy bin,
$P_{N,i}$, $P_{E,i}$ and $P_{S,i}$ are the probability density functions (PDFs) for NR, ER and SA,
respectively, and 
$f_{N,i}$ and $f_{S,i}$ are the NR and SA event fractions
in the $i^{\rm th}$ energy bin, respectively.

\begin{figure}[!htp]
\centering
\centerline{\scalebox{1.0}{\includegraphics[width=0.5\textwidth]{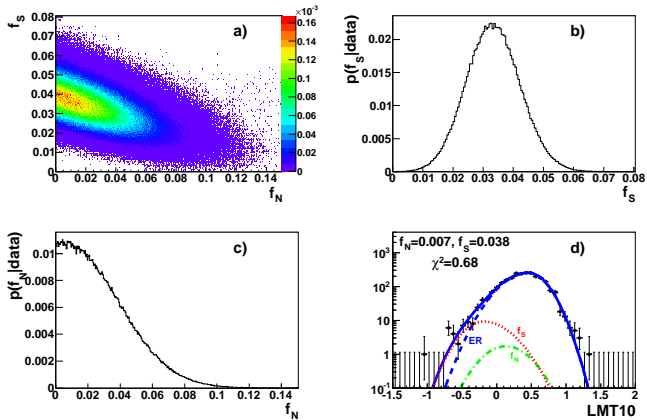}}}
\caption{{\bf a)} Two-dimensional plot of $f_{N,i}$ (horizontal) versus $f_{S,i}$ (vertical). {\bf b)} Projections of  
$f_{S,i}$ and {\bf c)} $f_{N,i}$ for the 6~keV bin of detector 9. {\bf d)} The fitted LMT10 distribution
for the 6~keV bin of detector 9.
\vspace{-15pt}
}
\label{fig:bat_f0Nf1}
\end{figure}

We determine $f_{N,i}$ and $f_{S,i}$ using the Bayesian Analysis Tool (BAT) program~\cite{paper_BAT} 
with prior PDFs for $f_{N,i}$ and $f_{S,i}$ that are flat between 0 and 1.
The Bayesian analysis produces posterior PDFs that reflect the degree of preference
for the parameter values based on the experimental data.
As an example, Fig.~\ref{fig:bat_f0Nf1}(a) shows contours of the two-dimensional posterior PDFs
for $f_{N,i}$ and $f_{S,i}$ for the 6~keV energy bin in one of the detector modules. 
The posterior PDF of each parameter is the projection of the two-dimensional PDF onto
that parameter's axis as shown in Figs.~\ref{fig:bat_f0Nf1}(b) and (c). 
To test the sensitivity to the choice of priors, we repeated this analysis
with the flat prior PDF replaced with a Jeffrey's prior~\cite{paper_jprior}. 
The Jeffrey's prior results agree with those with the flat prior within $1\sim2$ percent. 

\begin{figure}[!hbp]
\vspace{-5pt}
\centerline{\scalebox{1.0}{\includegraphics[width=0.45\textwidth]{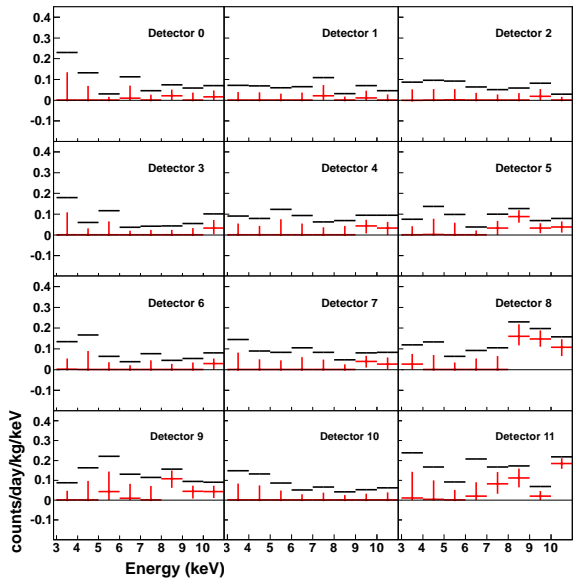}}}
\caption{
Nuclear recoil event rates for all detectors. The black horizontal bar indicate 90\% C.L. upper limits, the red vertical lines
denote the 68\% C.L. interval, and the red horizontal bars the most probable values.
\vspace{-5pt}
} \label{fig:bat_NR_det1}
\end{figure}

The NR event rate for each detector, determined
from $f_{N,i}$ and the event selection efficiency, 
is shown in Fig.~\ref{fig:bat_NR_det1}. 
The black horizontal bars indicate the 90\% confidence level (C.L.) upper limits and the red vertical lines
denote the 68\% C.L. intervals.  The red horizontal lines mark the most probable NR rate values.
\begin{figure}[!htp]
\includegraphics[width=0.36\textwidth]{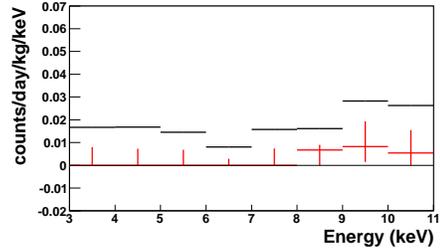}
\caption{
Total nuclear recoil event rates from the combined results from nine detectors (without detector 0, 8 and 11). 
\vspace{-15pt}
} \label{fig:bat_NR_combined_2}
\end{figure}

\begin{table}[!hbp]
\vspace{-5pt}
\begin{center}
\begin{tabular}{|c|c|c|c|}
\hline
\multicolumn{4}{|c|}{SA contamination level} \\
\multicolumn{4}{|c|}{(counts/day/kg/keV)} \\
\hline
 Detector & Level & Detector & Level \\
\hline
 0  &0.203 $ \pm$ 0.026&6  &0.087 $ \pm$ 0.021\\
 1  &0.071 $ \pm$ 0.017&7  &0.076 $ \pm$ 0.025\\
 2  &0.066 $ \pm$ 0.020&8  &0.238 $ \pm$ 0.025\\
 3  &0.089 $ \pm$ 0.024&9  &0.123 $ \pm$ 0.025\\
 4  &0.039 $ \pm$ 0.020&10 &0.014 $ \pm$ 0.026\\
 5  &0.064 $ \pm$ 0.018&11 &0.205 $ \pm$ 0.024\\
\hline
\end{tabular}
\caption{SA contamination level for each detector averaged over the 3-10 keV energy bins.
\vspace{-15pt}
}
\label{tab:SA}
\end{center}
\end{table}

The rates of SA events in the 3-11 keV energy range averaged over the detector modules are shown in Table~\ref{tab:SA}. 
The SA background levels of detectors 0, 8 and 11 are about three times as high as 
the average of the remaining detectors. For this reason, these three detectors are excluded from the average NR event rate determination. 
This reduces the total exposure used in the final analysis to 24524.3 kg$ \cdot $days.
Figure~\ref{fig:bat_NR_combined_2} shows the 68\% C.L. intervals and 90\% C.L. upper limits on the NR event rate from the combined PDF of the remaining nine detector modules.
These limits include systematic effects from uncertainties in the LMT10 PDFs for each event type
and their crystal-to-crystal deviations. No significant excess of NR events is observed. 

\begin{figure*}[!htp]

\begin{minipage}[t]{75mm}
\includegraphics[width=1.0\textwidth]{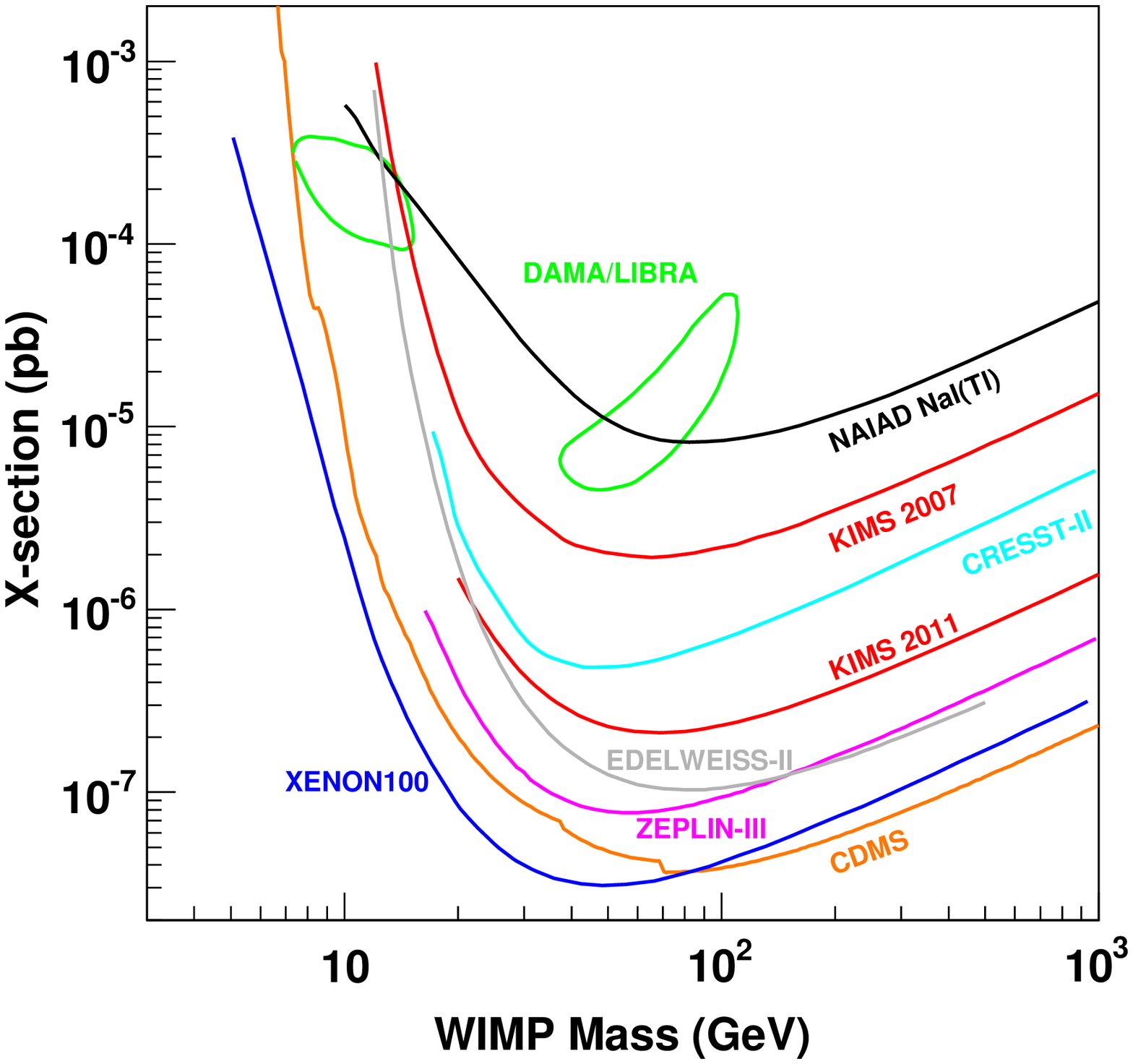}
\end{minipage}
\begin{minipage}[t]{75mm}
\includegraphics[width=1.0\textwidth]{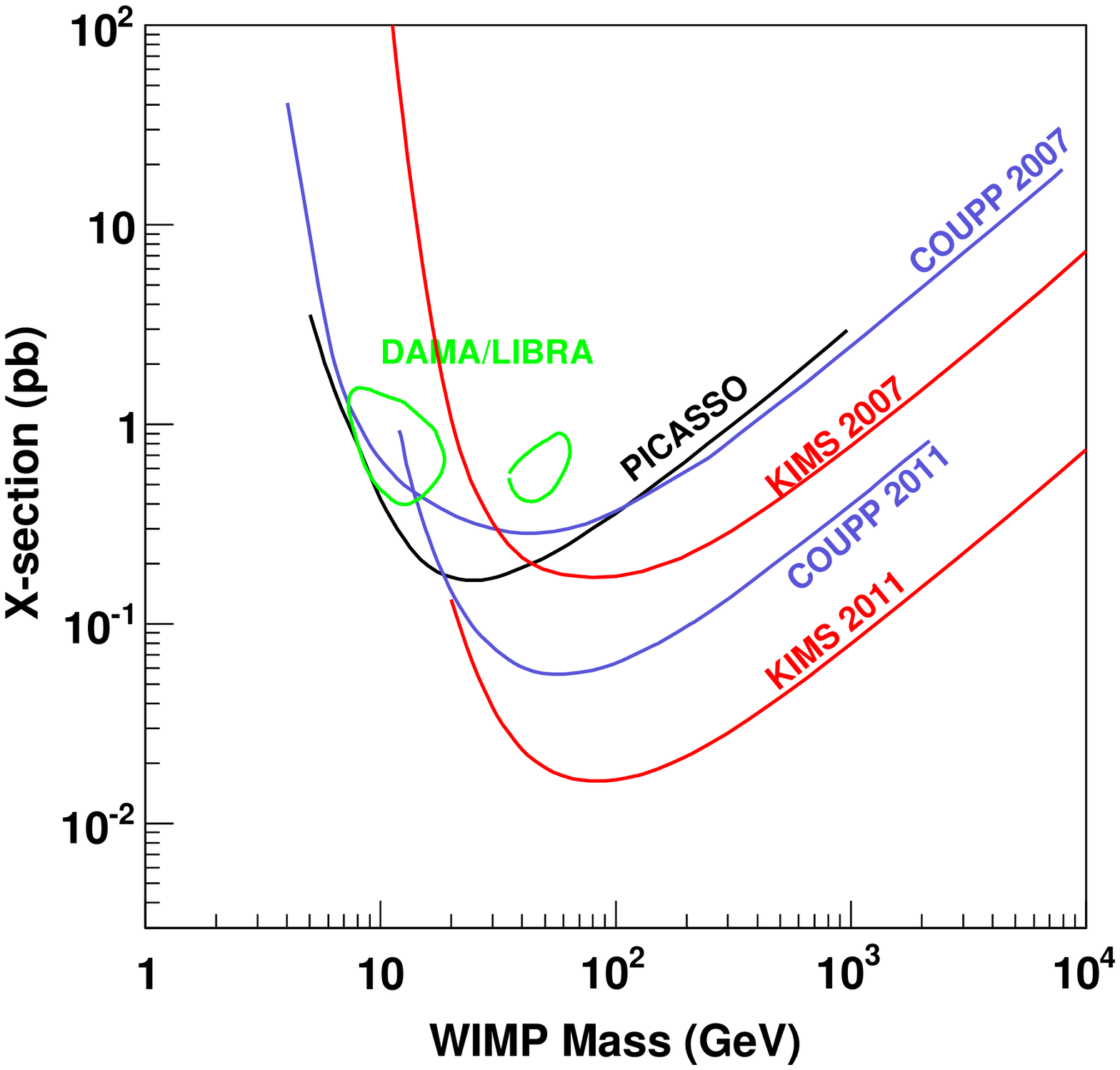}
\end{minipage}
\caption{The 90 \% exclusion limits on {\bf (Left)} SI 
WIMP-nucleon and {\bf (Right)} SD WIMP-proton cross sections.
In both plots DAMA results interpreted by Savage \etal ~\cite{savage} are used (3$\sigma$ contours are drawn).
The SI plot includes
NAIAD~\cite{ana_naiad2005}, CRESST-II~\cite{cresst2}, EDELWEISS-II~\cite{edelweiss2}, ZEPLIN-III~\cite{xeplin3},
XENON100~\cite{xenon100} and CDMS~\cite{cdms2} limits. The SD plot includes
PICASSO~\cite{picasso} and COUPP~\cite{coupp} limits.
\vspace{-15pt}
} \label{fig:SI}
\end{figure*}

Assuming the standard halo model~\cite{paper_Lewin}, we translate these
measurements to the cross section limits for WIMP-nucleon
SI interactions and WIMP-proton SD interactions that are presented 
in the left and right panels of Fig.~\ref{fig:SI}, respectively.  
The limits shown in the two figures are about one order-of-magnitude more stringent than the
previous KIMS results~\cite{prllee}, due primarily
to the identification of the SA background component and the larger exposure. 
The WIMP-proton SD interaction cross section limits are the most stringent to date. 

\begin{figure}[!htp]
\vspace{-5pt}
\includegraphics[width=0.4\textwidth]{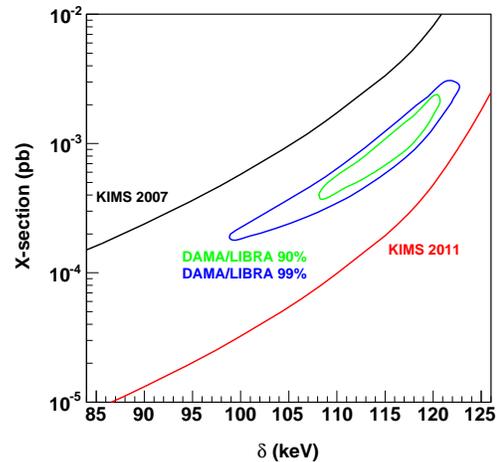}
\caption{
The allowed parameter space for DAMA/LIBRA~\cite{idm_chang} and the limits reported here for a 70~GeV WIMP mass in iDM model.
$\delta$ is the mass split between the ground and excited states of the WIMP. 
The astronomical parameters from Ref.~\cite{idm_chang} are used. 
\vspace{-20pt}
} \label{fig:idm_70GeV}
\end{figure}

The NR event rate limits also have important implications for the interpretation of
the DAMA annual modulation signal, which has an
amplitude of $0.0183 \pm 0.0022$~counts/day/kg/keV in the 2-4~keV energy range 
in \naitl scintillators~\cite{paper_dama_libra}.  Considering the different quenching factors of Iodine for
\naitl and \csitl~\cite{skkim10,Qf_CsI}, the 2-4~keV DAMA energy range corresponds to 3.6-5.8~keV in KIMS, which
is included in the first three bins in Fig.~\ref{fig:bat_NR_combined_2}.
Our 90\% C.L. upper limit on the NR event rate in the 3.6-5.8~keV energy range is 0.0098~counts/day/kg/keV,
which is well below the DAMA signal amplitude. 
Therefore, any scenario involving Iodine as the target, such as the iDM model, 
is incompatible with our limits.  
As an example, the parameter space allowed for DAMA in the iDM model and our
exclusion limits for a WIMP of mass 70~GeV are presented in Fig.~\ref{fig:idm_70GeV}.  
An alternative iDM interpretation considers Thallium,
which is present at the $10^{-3}$ level in both the DAMA and KIMs detectors~\cite{chang_2011},
as the dominant target, can be addressed by our results.
We estimate the quenching factors for Thallium
in \naitl and \csitl using a semi-empirical calculation~\cite{tretyak} 
and find $ \frac{Q_{\mathrm{CsI}}^{\mathrm{I}}}{Q_{\mathrm{NaI}}^{\mathrm{I}}} \approx \frac{Q_{\mathrm{CsI}}^{\mathrm{Tl}}}{Q_{\mathrm{NaI}}^{\mathrm{Tl}}} $, 
where $Q_{\mathrm{CsI,NaI}}^{\mathrm{I,Tl}}$ is the quenching factors of \csitl and \naitl for iodine and thallium ions.  This
indicates that the corresponding energy range in KIMS for Thallium is about the same as that for Iodine.
Therefore our conclusion does not change when Thallium is considered as the dominant target.

In conclusion, we report improved limits for WIMP-nucleon cross sections using a data sample collected
with a 103.4~kg \csitl scintillator detector array with a total exposure of 24524.3 kg$ \cdot $days.
We identified and characterized a low energy background due to a contamination
of alpha emitters on the surfaces of the crystals and incorporated it
into the PSD analysis. 
No significant signals for NR events are observed and we determine 90\% C.L. upper limits on NR event rates,
and improved limits on WIMP-nucleon cross sections, including the most stringent limits to date on 
WIMP-proton SD scattering. 
The NR event rate upper limit is below the DAMA/LIBRA annual modulation
amplitude in the corresponding energy region, disfavoring iDM model interpretations. 

%\acknowledgments
 We are very grateful to the Korea Midland Power Co. and Korea Hydro and Nuclear Power Co.  for
 providing the underground laboratory space at Yangyang.
 This research was supported by the WCU program (R32-10155) and Basic Science Research Grant (KRF-2007-313-C00155) through the National Research Foundation
 funded by the Ministry of Education, Science and Technology of Korea.

\end{document}